\documentclass[twoside,english,nofootinbib,preprintnumbers,aps,11pt]{revtex4}

\usepackage{euscript,amssymb}
\usepackage{amsthm,latexsym}
\usepackage{graphicx}
\usepackage{amsfonts}
\usepackage{amsmath}
\usepackage{amssymb}
\usepackage{fancyhdr}
\usepackage{epsfig}
\usepackage{color}
\usepackage{esint}
\usepackage{soul}
\usepackage{afterpage}
\usepackage[papersize={8.5in,11in}]{geometry}

\usepackage[normalem]{ulem}
\usepackage{color}

\newcommand{\be}{\begin{equation}}
\newcommand{\ee}{\end{equation}}

\begin{document}


\title{The axial symmetry of Kerr without the rigidity theorem}


\author{Jerzy Lewandowski}

\email[]{Jerzy.Lewandowski@fuw.edu.pl}

\author{Adam Szereszewski}

\email[]{Adam.Szereszewski@fuw.edu.pl}

 \affiliation{\vspace{6pt} Faculty of Physics, University of Warsaw, Pasteura 5, 02-093 Warsaw, Poland\\}

\begin{abstract} Local condition that imply the no-hair property of black holes  are completed. The conditions take the form of  constraints on the geometry  of the $2$-dimensional crossover surface of  black hole horizon. They imply also the axial symmetry without the rigidity theorem. This is the new result contained in this letter.  
The family  of the solutions to our constraints is 2-dimensional and can be  parametrized  by the area and angular momentum.  The constraints are induced by our assumption  that the horizon is of the Petrov type D.  Our result applies to all the bifurcated Killing horizons: inner/outer black hole horizons as well as  cosmological horizons.  Vacuum spacetimes with a given cosmological constant can be reconstructed from our solutions via Racz's black hole holograph.         
 \end{abstract}

\date{\today}

\pacs{???}

\maketitle

\section{Introduction} The family of vacuum stationary  asymptotically
flat black holes   is two dimensional and consists of the Kerr
solutions \cite{Kerr,Exact}. The metric tensor of each of the
spacetimes in this family  is  uniquely characterized by  two
parameters. The standard parameters are the known "$m$" and "$a$", but
instead, they may be chosen to be invariants of the geometry of the
Kerr event horizon, like the area of a $2$-dimensional cross-section
and the angular momentum.   A mathematical proof of that result uses
the global properties of those spacetimes \cite{Hawking-Ellis}. A key
step in the proof is the rigidity theorem. It  states the existence of
the second Killing vector, a generator of  the axial symmetry.  The
analysis of the symmetry reduced vacuum  Einstein equations completes
the proof \cite{heusler}. The spacetime characterization of the Kerr
metric is available in the literature and it provides the uniqueness properties \cite{Mars1,Mars2}.   

An alternative approach consists in a local rather than global characterization of black hole spacetimes.   And then, how to distinguish  exactly the same  two dimensional family of the Kerr spacetimes imposing some conditions locally?  An answer was found in  the theory of isolated horizons \cite{ABL,LPkerr,DLP1,DLP2}.  Isolated horizon is a local  generalization of  event horizon of a stationary black hole spacetime (in fact quasi-local: cross-section is assumed to be a compact manifold). Its intrinsic geometry does have local degrees of freedom even after solving the constraints 
that follow from the vacuum Einstein equations.  The degrees of freedom are a Riemannian metric tensor $g_{AB}$ and a  rotation $1$-form 
potential $\omega_A$ defined on a $2$-dimensional topological sphere $S$, a cross-section of isolated horizon \cite{ABL}. 
But what condition can reduce the freedom of an arbitrary metric tensor to a single number, the total area $A$, and an arbitrary  $1$-form filed  to a single number, the total 
angular momentum $J$?  An assumption that the spacetime Weyl tensor is Lie dragged along the isolated horizon by its null symmetry generator and that the Petrov type of the Weyl tensor at the isolated horizon  is D amounts to  a system of two $4$th order partial  differential equations     on $g_{AB}$ and $\omega_A$, referred to as the type D equation \cite{LPkerr,DLP1}.  The equation was derived and all the  axisymmetric solutions were found  assuming the vacuum Einstein equations with arbitrary  cosmological constant \cite{LPkerr,DLP1,DLP2}.  As desired, the family of the axisymmetric  solutions is $2$-dimensional, each solution is determined by values the pair $(A,J)$ takes.  However, why should the isolated horizon  be axisymmetric?    The answer to that question is the new result we present in the current letter.  The clue  is to consider a bifurcated isolated horizon,  that is Racz's  framework of the black hole holograph \cite{Racz1,Racz2}. A similar result was obtained by
Racz and collaborators by using the characteristic initial value problem for the Killing spinors \cite{Racz_private_communication}. It motivated us to derive the 
existence of  the axial  symmetry directly from the type D equations \cite{DLP1}.  
      
\section{The type D equation} 
Consider a $4$-dimensional spacetime $M$ endowed with a metric tensor $g_{\mu\nu}$. 
The quasi-local  generalization of a black hole we will be concerned with in this paper is a $3$-dimensional null surface 
$$H\ \subset\ M ,$$   
such that there exists a vector field $\xi$ in a neighborhood of $H$,  that is tangent to and null at  $H$ and 
Lie drags along $H$ the following spacetime structures:
\begin{equation}\label{LxignablaR}
{\cal L}_\xi g_{\mu\nu}{}_{|_{H}}\ =\  [{\cal L}_\xi, \nabla_{\mu}]{}_{|_{H}}\ =\ {\cal L}_\xi R_{\mu\nu\alpha\beta}{}_{|_{H}}\ =\ 0 , 
\end{equation}  
where  $\nabla_\mu$ and  $R_{\mu\nu\alpha\beta}$ are the spacetime covariant derivative and its Riemann tensor, respectively.  
The null surface $H$ with these properties will be called the stationary to the second order horizon.
We  assume  that this null symmetry acts non-trivially at each null geodesic in $H$. Suppose,  that the vacuum Einstein equations 
\begin{equation}\label{Ee}
G_{\mu\nu}\ +\ \Lambda g_{\mu\nu}\ =\ 0  
\end{equation}   
hold  at $H$, as well as their  transversal derivative. Topologically, $H$ is a cylinder
$$ H\ =\ S\times \mathbb{R} $$    
where $\mathbb{R}$ corresponds to the null geodesics.   

A consequence of those assumptions is, 
\begin{equation}
\xi^\mu\nabla_\mu \xi^\nu{}_{|_{H}}\ =\ \kappa \xi^\nu, \ \ \ \ \ \ \ \ \ \ \ \ \ \ \ \kappa\ =\ {\rm const} . 
\end{equation}  
The second equality is the zeroth law of non-expanding null surfaces thermodynamics \cite{ABL}.    
We are assuming throughout this paper, that the horizon is non-extremal, that is
\begin{equation}\label{non-ext}
\kappa\  \not= \ 0 . 
\end{equation} 
   
The constraints implied on the geometry at $H$ by (\ref{Ee},\ref{LxignablaR},\ref{non-ext}), allow to determine
all the components of $g_{\mu\nu}$, $\nabla_\mu$ and $R_{\mu\nu\alpha\beta}$ at $H$, by 
data 
\begin{equation}\label{data}(g_{AB},\omega_A)\end{equation} 
defined on a spacelike $2$-dimensional cross-section of $H$ diffeomorphic to  $S$,  namely, the induced metric tensor $g_{AB}$ 
and the pullback $\omega_A$ of the rotation $1$-form potential 
$\omega^{(\xi)}_a$ defined on $H$ by the following equality
\begin{equation}\label{omega}
\nabla_a \xi^b{}_{|_{H}}\ =\ \omega^{(\xi)}_a\xi^b,
\end{equation} 
wherever 
$$\xi \neq 0$$
(the lower case lattice indices $a,b,...$ correspond to the bundle tangent to $H$). 
The metric tensor $g_{AB}$ is independent of a choice of the cross-section  (we identify two different cross-sections
in the natural way, using their intersections with the null geodesics contained in $H$). 
On the other hand, the $1$-form $\omega_A$ depends on the choice of a cross-section, and transforms as follows
\begin{equation}\label{omega'}
\omega'_A \ =\ \omega_A\ +\ \kappa f_{,A} .
\end{equation}        
Actually,  there is a distinguished   section $S$ of $H$, at which the vector field $\xi$
vanishes,
\begin{equation}\label{S}
\xi{}_{|_{S}}\ =\ 0 ,
\end{equation}  
provided $H$ is sufficiently complete. We call it a crossover section. However, the $1$-form $\omega_A$ can be defined on $S$ via
a limit only, and the limit is unique up to the transformations (\ref{omega'}).  Indeed, to deal with that section 
we introduce at $H$ a nowhere vanishing nonsingular  vector field $\ell$ and a corresponding function $u$
such that
\begin{equation}\label{ell0}
 \ell^a\nabla_a\ell = 0, \ \ \ \ \ \ \ \ \ \ \ \ell^au_{,a}\ =\ 1, 
\end{equation} 
and  
\begin{equation}\label{ell}
\xi{}_{|_{H}}\ =\ \kappa u\ell.  
\end{equation} 
Then the limit of $\omega_A$ obtained by using the slices
$$u={\rm u_0},$$ 
and $u_0\rightarrow 0$ is just the pullback on $S$ of  $\omega^{(\ell)}_a$ defined in the analogous way to (\ref{omega}).  
However still, the choice of $u$ is ambiguous, and the ambiguity corresponds exactly to (\ref{omega'}). 

The data $(g_{AB}, \omega_A)$ defined on $S$  
is free, in the sense that for every possibility there exists a stationary to the second order
horizon embedded in a vacuum spacetime. The Gaussian curvature $K$ is a scalar invariant of the data. The second one, a pseudo 
scalar $\Omega$ is defined by the rotation $2$-form
\begin{equation}\label{Omega}
2\partial_{[A}\omega_{B]} \ =:\ \Omega\, \eta_{AB} ,
\end{equation} 
where $\eta_{AB}$ is the area $2$-form. 

Since the data (\ref{data}) determines the spacetime Weyl tensor at $H$, we attribute to $(g_{AB},\omega_A)$   the Petrov type         
of the Weyl tensor at $H$ \cite{DLP1}. It turns out, that in a generic case   $H$ is of the Petrov type II, except for the section 
$S$  (\ref{S}) where the Petrov type is D. All the $H$ is of the Petrov type D,  the same as  the Kerr 
horizon, if and only if  some equation is satisfied by the data $(g_{AB}, \omega_A)$. To write it, it is convenient to use a complex 
null frame $(m^A, \bar{m}^A)$ tangent to $S$ such that
\begin{equation}
g_{AB}\ =\ m_A\bar{m}_B\ +\  m_B\bar{m}_A , \ \ \ \ \ \ \ \ \ \ \ \ \ \eta_{AB} \ =\ i\left(\bar{m}_A{m}_B - \bar{m}_B{m}_A\right) .
\end{equation}
The Petrov type D equation reads \cite{DLP1} ($D_A$ below, is the metric, torsion free covariant derivative on $S$ corresponding to $g_{AB}$)
\begin{equation}\label{nablanabla}
\bar{m}^A\bar{m}^BD_AD_B\left(K-\frac{\Lambda}{3}+i\Omega\right)^{-\frac{1}{3}}\ =\ 0,
\end{equation} 
where a necessary condition is
\begin{equation}
K-\frac{\Lambda}{3}+i\Omega\ \not= \ 0 .
\end{equation}

To write down that equation explicitly, let us use local complex coordinates $(z,\bar{z})$ on $S$
and write the metric tensor as
\begin{equation}
g_{AB}dx^A dx^B\ =\ \frac{2}{P^2} dz d\bar{z} , 
\end{equation}  
and a rotation $1$-form potential 
\begin{equation}
\omega_A dx^A\ =\ \omega_zd{z} + \omega_{\bar z}d{\bar z} . 
\end{equation}  
The complex frame becomes
\begin{equation}
m^A\partial_A\ =\ P\partial_z, \ \ \ \ \ \ \ \ \ \ \ \ \ \ \bar{m}_Adx^A \ =\ \frac{1}{P}dz  . 
\end{equation}
The volume form is
$$ \frac{1}{2}\eta_{AB}dx^A \wedge dx^B\ =\ i\frac{1}{P^2}dz\wedge d{\bar z} . $$
The differential operator (\ref{nablanabla}) becomes
\begin{equation}
\bar{m}^A\bar{m}^BD_AD_B\  =\  \partial_{\bar{z}}\circ P^2 \circ \partial_{\bar{z}} .
\end{equation} 
That second order equation (\ref{nablanabla}) can be easily integrated once, and the result is
\begin{equation}\label{Czbar}
\partial_{\bar{z}}\left(K-\frac{\Lambda}{3}+i\Omega\right)^{-\frac{1}{3}} \ =\ \frac{F}{P^2} ,\ \ \ \ \ \ \ \ \ \ \ \ \partial_{\bar{z}}F\ =\ 0.
\end{equation} 
We will use it in the next section. 

The equation (\ref{nablanabla}) can be solved explicitly in  the axisymmetric case on a sphere \cite{LPkerr,DLP2}. The solutions set a 
$2$-dimensional family labelled by the area and angular momentum. In the case of 
$$\Lambda=0,$$ 
$A$ and $J$ range $\mathbb{R}^+\times \mathbb{R}$ and each of them corresponds 
to either inner,  or outer, respectively,  horizon in the non-extremal Kerr spacetime (including Schwarzschild), or to the horizon in the spacetime  
obtained from the extremal Kerr metric via the near extremal horizon limit  \cite{Horowitz,LivRevNHG} when
$$A=8\pi J.$$ 
Similar embeddings exist in the   $\Lambda \neq 0$ case \cite{DLP2} (just the extremity conditions are  more complicated).  
However, possibly, there are  also non-axisymmetric solutions of the type D equation (\ref{nablanabla}).  What implies the axial symmetry?


\section{Bifurcarted type D horizon}   
A bifurcated stationary to the second order horizon consists of two intersecting stationary to the second order horizons, 
say $H$ and $H'$. So we have now the symmetry $\xi$ (\ref{LxignablaR}) and analogous symmetry $\xi'$ 
for $H'$. Usually that is a same vector field $\xi=\xi'$ that vanishes at their intersection, so let us assume that is what happens.
Hence the intersection  
$$ S\ =\ H\cap H' $$
is the considered above  crossover section  of $H$ and a cross-section of $H'$ at the same time.  Now,  in addition to the data
$(g_{AB}, \omega_A)$ (\ref{data}) induced on $S$ by the intrinsic geometry of $H$, there is new data $(g'_{AB}, \omega'_A)$
induced by an intrinsic geometry of $H'$. Obviously,
$$ g'_{AB}\ =\ g_{AB}$$
because they are induced by a same spacetime metric tensor
$g_{\mu\nu}$.  On the other hand, if on $H'$ we introduce 
the decomposition (\ref{ell}) and choose $\ell'$ such that
\begin{equation}
\ell^\mu\ell'_{\mu}{}_{|_S}\ =\ -1,
\end{equation}          
then
\begin{equation}
\omega_A\ =\  -\ell'_\mu\nabla_A\ell^\mu\ =\ \ell_\mu\nabla_A\ell'^\mu\ =\ -\omega'_A .
\end{equation}
Hence, the corresponding  invariants $K'$ and $\Omega'$ are 
\begin{equation}
K'\ =\ K, \ \ \ \ \ \ \ \ \ \ \ \ \Omega'\ =\ -\Omega . 
\end{equation}
Suppose now, that the spacetime Weyl tensor is of the type D on the entire bifurcated horizon $H\cup H'$. 
That implies two type D equations, one on the data $(g_{AB}, \omega_A)$ and  $(g_{AB}, -\omega_A)$,
respectively, namely
\begin{align}
\bar{m}^A\bar{m}^BD_AD_B\left(K-\frac{\Lambda}{3}+i\Omega\right)^{-\frac{1}{3}}\ &=\ 0,\label{nablanabla'}\\
\bar{m}^A\bar{m}^BD_AD_B\left(K-\frac{\Lambda}{3}-i\Omega\right)^{-\frac{1}{3}}\ &=\ 0.
\end{align}

But the second  equation is equivalent to 
\begin{equation}\label{nablanablaconj}
{m}^A{m}^BD_AD_B\left(K-\frac{\Lambda}{3}+i\Omega\right)^{-\frac{1}{3}}\ =\ 0
\end{equation}
that may be considered a conjugate type D equation, on the data $(g_{AB}, \omega_A)$.
In conclusion, now given data $(g_{AB},\omega_A)$ is subject to two equations, the type D equation (\ref{nablanabla'})
and the conjugate type D equation (\ref{nablanablaconj}). 

We can use  again the partial solution method that led to (\ref{Czbar}) and apply now it to (\ref{nablanablaconj}). 
In that way we obtain two equations, namely
\begin{align}\label{Cz}
\partial_{\bar{z}}\left(K-\frac{\Lambda}{3}+i\Omega\right)^{-\frac{1}{3}} \  &=\ \frac{F}{P^2},\\
\partial_{{z}}\left(K-\frac{\Lambda}{3}+i\Omega\right)^{-\frac{1}{3}}  \ &=\ \frac{\bar{G}}{P^2}\nonumber\, 
\end{align}
with the additional conditions on $F$ and $G$ ,
\begin{equation}\label{Fbarz}     
\partial_{\bar z} F\ =\ 0\ =\ \partial_{\bar z} G .
\end{equation} 
The standard integrability condition following from the symmetry of the second derivatives gives
\begin{equation}\label{Czbarz}
\left( \frac{F}{P^2} \right)_{,z}\ -\  \left( \frac{\bar{G}}{P^2} \right)_{,\bar{z}}\ =\ 0 .
\end{equation}
And now surprise: consider the (a priori complex valued) vector field 
 \begin{equation}\label{Phi}
\Phi\ =\ {F}\partial_{z}\ -\  {\bar{G}}\partial_{\bar{z}};
\end{equation}
the equations (\ref{Fbarz},\ref{Czbarz})  are just the three components of the Killing equation
\begin{equation}\label{LPhig}
{\cal L}_{\Phi} g_{AB}\ =\ 0 .
\end{equation}
Therefore both, the real and imaginary parts of $\Phi$ are the Killing vectors of $g_{AB}$. 
The vector field $\Phi$ has a clear geometric definition, namely \cite{LPkerr}
\begin{equation} \label{Phi'}     
\Phi^A\eta_{AB}\ =\ i \partial_B\, \left(K-\frac{\Lambda}{3}+i\Omega\right)^{-\frac{1}{3}}.
\end{equation}
This definition is global, in the sense that both, $\Phi$ and $\eta_{AB}$ are is defined on  $S$. Another consequence is
\begin{equation}\label{32}
{\Phi}^A (K+i\Omega)_{,A}\ =\ 0 .
\end{equation}
Since the Gaussian scalar  $K$ is annihilated  by the (even complex) Killing vector  $\Phi$ itself,
and so is the area two form $\eta_{AB}$, (\ref{32}) implies also

\begin{equation}\label{LPhiOmega}
{\cal L}_{\Phi} d\omega\ =\ {\cal L}_{\Phi}\left(\frac{1}{2}\Omega\,\eta_{AB} dx^A\wedge dx^B\right)\ =\ 0 .
\end{equation} 
The vector field $\Phi$ may  in general be complex valued, but the symmetry equations (\ref{LPhig},\ref{LPhiOmega})
are satisfied independently by the real and imaginary parts  Re$\Phi$, Im$\Phi$, respectively.  If 
any of those vectors   does not vanish at any point $x\in S$, then it generates a  symmetry group of the data $(g_{AB}, \omega_A)$
modulo (\ref{omega'}) provided $S$ is  connected (if it is not, we may repeat our argument for each connected component).    On the other hand, if
$$\Phi\ = \ 0$$
identically on $S$, then both, $K$ and $\Omega$ are constant, the the constancy of $K$ implies that 
that $g_{AB}$ has Killing vectors that annihilate the area element $\eta_{AB}$, hence finally they also
Lie drag $d\omega$. 

In summary,  every data $(g_{AB},\omega_A)$ modulo (\ref{omega'}) that is a solution to  both, the type D equation and the conjugate  type D equation
(\ref{nablanabla}, \ref{nablanablaconj}), has (at least)  a $1$-dimensional infinitesimal symmetry.  If $S$ is a $2$ sphere, then more can be concluded. 
Then, vector field $\Phi$ generates  the group O$(2)$, hence all  the solutions are axisymmetric. For the axi-symmetric  solutions of the type D equation (\ref{nablanabla})  the local no-hair  theorem applies  \cite{LPkerr,DLP2}. The family of the solutions is $2$-dimensional, and can be parametrized by the area and angular momentum of the horizon.  The space of the axisymmetric solutions of the type D equation (\ref{nablanabla})  admits the symmetry  
$$\Omega\mapsto -\Omega,$$  
therefore each of those solutions is simultaneously a solution to the conjugate equation (\ref{nablanablaconj}).


\section{Summary and discussion} 
The best way to summarize  our result is to do it from the point of view of the black hole holograph \cite{Racz1,Racz2}.  
Every  pair $(g_{AB},\omega_A)$, that is a metric tensor and a $1$-form  respectively, defined on a $2$-dimensional 
sphere $S$ uniquely determine a solution to the vacuum Einstein equations (with possibly non-zero cosmological constant) 
such that  $S$ is the cross-over surface of a bifurcated Killing horizon and the  the solution is defined in the future and the 
past, respectively,  wedge of spacetime. That data is free. A local condition that restricts that data considerably is 
assumption that the data defines a type D horizon. The corresponding equations on $(g_{AB},\omega_A)$, the type D one 
and the conjugate type D one, respectively, imply the axial symmetry.  In that way the axial symmetry emerges from within,
without the rigidity mechanism. The space of solutions is two dimensional, can be uniquely parametrized by the area $A$
and angular momentum $J$. In $\Lambda=0$ case, each of those $(g_{AB},\omega_A)$ gives rise via the black hole holograph 
construction to a Kerr spacetime unless the values of   $A$ and $J$ correspond to the extremal case. In that case the black hole
holograph produces from $(g_{AB},\omega_A)$ a near horizon geometry \cite{LivRevNHG}.   In all the cases the two-wedge
spacetime metric is analytic and a complete spacetime can be obtained as a maximal analytic extension. Similarly, in the
$\Lambda\not=0$ case, the black hole holograph produces either the corresponding Kerr - (anti) de Sitter spacetime    
or near horizon geometry, however in that case some of the horizons do not prevent emergence of naked singularity.         

\bigskip
\bigskip

\noindent{\bf Acknowledgments.}   The idea to consider bifurcated isolated horizons  that was the key new element in our work
on the type D horizons, and new motivation to look for the axial symmetry   has come  from Istvan Racz. We appreciate it very much.   

This work was partially supported by the Polish National
Science Centre\\ grant No. 2015/17/B/ST2/02871 .

\end{document}